\documentclass[letter]{aa} 

\usepackage{graphicx}
\usepackage{txfonts}
\begin{document}

   \title{Extreme Starlight Polarization in a Region with Highly Polarized Dust Emission}
   \titlerunning{Extreme starlight polarization in a high $p_{353}$ region}


   \author{Georgia V. Panopoulou
          \inst{1}
          \and
          Brandon S. Hensley\inst{2}
          \and
          Raphael Skalidis\inst{3,4}
          \and
          Dmitry Blinov\inst{3,4,5}
          \and
          Konstantinos Tassis\inst{3,4}
          }

   \institute{California Institute of Technology, MC249-17, 1200 East California Boulevard, Pasadena, CA 91125, USA\\
              \email{panopg@caltech.edu}
         \and
             Spitzer Fellow, Department of Astrophysical Sciences, Princeton University, Princeton, NJ 08544, USA
        \and
        Department of Physics and ITCP, University of Crete, Voutes, 70013 Heraklion, Greece
        \and
        Institute of Astrophysics, Foundation for Research and Technology-Hellas, Voutes, 70013 Heraklion, Greece
        \and
        Astronomical Institute, St. Petersburg State University, Universitetsky pr. 28, Petrodvoretz, 198504 St. Petersburg, Russia
             }


 
  \abstract
   {Galactic dust emission is polarized at unexpectedly high levels, as revealed by \textit{Planck}.}
   {The origin of the observed $\simeq 20\%$ polarization fractions can be identified by characterizing the properties of optical starlight polarization in a region with maximally polarized dust emission.}
   {We measure the R-band linear polarization of 22 stars in a region with a submillimeter polarization fraction of $\simeq 20\%$. A subset of 6 stars is also measured in the B, V and I bands to investigate the wavelength dependence of polarization.}
   {We find that starlight is polarized at correspondingly high levels. Through multiband polarimetry we find that the high polarization fractions are unlikely to arise from unusual dust properties, such as enhanced grain alignment. Instead, a favorable magnetic field geometry is the most likely explanation, and is supported by observational probes of the magnetic field morphology. The observed starlight polarization exceeds the classical upper limit of $\left[p_V/E\left(B-V\right)\right]_{\rm max} = 9\%\,$mag$^{-1}$ and is at least as high as 13\%\,mag$^{-1}$ that was inferred from a joint analysis of {\it Planck} data, starlight polarization and reddening measurements. Thus, we confirm that the intrinsic polarizing ability of dust grains at optical wavelengths has long been underestimated.}
   {}

   \keywords{Polarization --
                ISM: magnetic fields --
                ISM: dust --
                submillimeter: ISM --
                 Galaxy: local interstellar matter
               }
   \maketitle
%

\section{Introduction} \label{sec:intro}

Sensitive submillimeter observations from the {\it Planck} satellite revealed surprisingly high levels of polarized emission from Galactic dust \citep{Planck_Int_XIX}. The observed polarization fractions in the {\it Planck} bands are higher than those anticipated by dust models consistent with starlight polarimetry \citep[e.g.,][]{Draine2009} and are challenging to reproduce with physical models \citep[e.g.,][]{Guillet2018}.

The same grains that radiate polarized emission in the infrared induce starlight polarization in the optical, and so the magnitudes of the two effects are tightly coupled. Indeed, the 353\,GHz polarization fraction $p_{353}$ and the V-band polarization per unit reddening $p_V/E(B-V)$ have a characteristic ratio of $\simeq$1.5\,mag \citep{Planck_Int_XXI,Planck_2018_XII}. However, applying this relation to the highest observed values of $p_{353} \simeq 20$\% implies starlight polarization far in excess of the classical upper limit of 9\%\,mag$^{-1}$ \citep{Hiltner1956, Serkowski1975}. By performing stellar polarimetry in a region with $p_{353} \simeq 20\%$, we seek to clarify the origin of the strong polarization.

If the derived $p_{353}$ is truly reflective of Galactic dust polarization and not, for instance, from over-subtraction of the Cosmic Infrared Background or Zodiacal Light \citep{Planck_2015_X, Planck_2018_XII}, then stars in this region should be highly polarized. Strong dust emission and starlight polarization could be the result of dust with unusual polarization properties, such as enhanced grain alignment, or simply of a favorable magnetic field geometry, in which case the magnetic field is nearly in the plane of the sky and has a uniform orientation along the line of sight. To discriminate among these possibilities, we consider whether the wavelength dependence of the interstellar polarization conforms to a typical ``Serkowski Law'' \citep{Serkowski1975} and whether the magnetic field orientation as probed by linear structures in HI emission \citep{Clark2018} is uniform along the line of sight.

In this Letter, we use RoboPol polarimetry of 22 stars within a small region to test whether stars in regions of high $p_{353}$ have $p_V/E(B-V) \simeq 13\%$\,mag$^{-1}$, as inferred by the analysis of \citet{Planck_2018_XII}.
We compare the derived $E(B-V)$ (appendix \ref{ssec:reddening}) to the observed optical fractional linear polarization (Section~\ref{ssec:pebv}). We find that the optical polarization is in excess of the classical upper limit with $p_V/E(B-V) \geq 13$\%\,mag$^{-1}$. The six stars with multi-band polarimetry show typical wavelength dependence of the polarized extinction (Section~\ref{subsec:serkowski}), while starlight polarization angles and HI emission data suggest a highly uniform magnetic field orientation (Section~\ref{ssec:angles}). We therefore argue that the observed $p_{353} \simeq 20\%$ arises from favorable magnetic field geometry and that the classical upper limit of $\left[p_V/E(B-V)\right]_{\rm max} = 9\%$\,mag$^{-1}$ underestimates the intrinsic polarizing efficiency of dust at optical wavelengths (Section \ref{sec:summary}).

\begin{figure*}[ht!]
\centering
\includegraphics[scale=1]{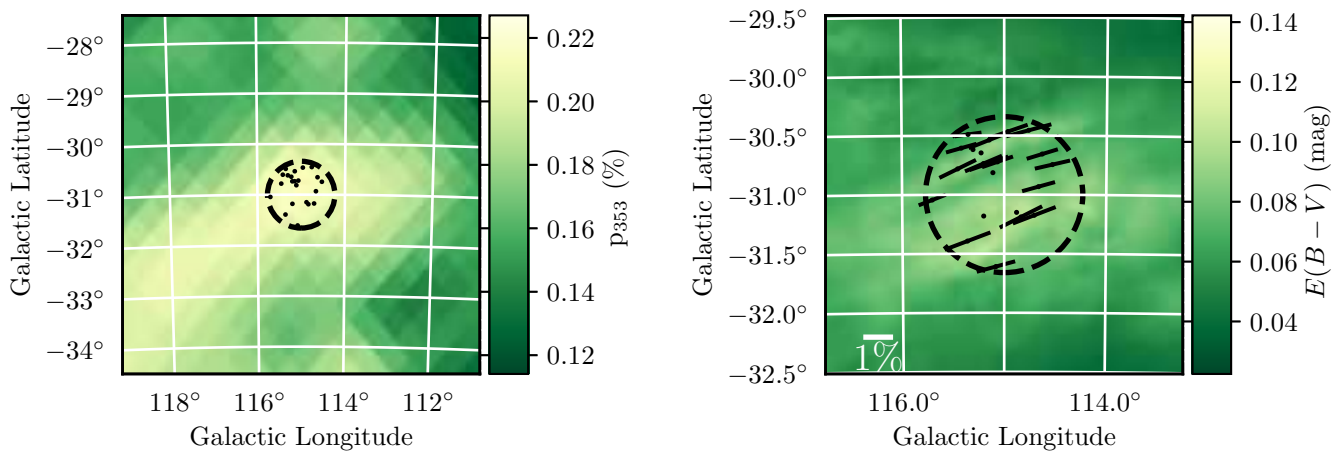}
\caption{Left: Map of fractional linear polarization of dust emission from \textit{Planck} at 353 GHz at a resolution of 80$\arcmin$ (GNILC). Right: The background image shows a map of $E(B-V)$ from \cite{PlanckGNILC2016} (PG). The map has a pixel size of 1.7\arcmin. For stars with a significant detection of $\hat{p}_{\rm R}$, we show segments that form an angle $\theta$ with the Galactic reference frame (according to the IAU convention). The length of each segment is proportional to the $\hat{p}_{\rm R}$ of the star (measured in the R band) and a line of $p=1\%$ is shown on the bottom left for scale. In both panels the dashed circle marks the region in which we have measured stellar polarization. The locations of stars in the observed sample are shown with black dots.}
\label{fig:planckp}
\end{figure*}

\section{Data} \label{sec:obs}

\subsection{RoboPol observations and Gaia distances}

Optical polarization observations were conducted using the RoboPol linear polarimeter \cite[][]{ramaprakash}. Stars were selected from the USNO-B1 catalog \citep{usnob} within a 80\arcmin-wide region centered on (l,b) = 115$^\circ$, -31$^\circ$ (circled region in Fig. \ref{fig:planckp}). We selected 22 stars brighter than 14\,mag in the R band with distances out to 1.5 kpc, based on the catalog of stellar distances derived from \textit{Gaia} \citep{bailer-jones}. The measurements were made in the Johnsons-Cousins R band. To investigate the wavelength dependence of polarization in the region, we observed 6 of these stars in three additional bands (B, V, I). All sources were placed at the center of the RoboPol field of view, where the instrumental systematic uncertainty is lowest.

Observations were conducted during 8 nights in the autumn of 2018. The unpolarized standard stars HD 212311, BD +32 3739, HD 14069, and G191-B2B, as well as the polarized star BD +59 389 \citep{schmidt} were observed during the same nights for calibration. In total we obtained 22 measurements of standard stars in the R band and 10 measurements in each of the B, V and I bands. We find systematic uncertainties on the fractional linear polarization of 0.16\% in the R band, 0.23\% in B, and 0.2\% in V and I. We follow the data calibration and reduction methods described in \cite{Panopoulou2019}.
 
The fractional linear polarization, $p$, and its uncertainty, $\sigma_p$, are calculated from the Stokes parameters $q$, $u$, taking into account both statistical and systematic uncertainties \citep[as in][]{Panopoulou2019}. Throughout the paper we use the modified asymptotic estimator proposed by \citet{plaszczynski} to correct for the noise bias on $p$. We refer to the noise-corrected $p$ as the debiased fractional linear polarization, $\hat{p}$, and use $\hat{p}_{\rm B}, \hat{p}_{\rm V}, \hat{p}_{\rm R}, \hat{p}_{\rm I}$ for measurements in different bands.

The polarization angle measured with respect to the International Celestial Reference Frame (ICRS): $\phi=0.5\arctan(u/q)$ is calculated using the two-argument arctangent function. The uncertainty in $\phi$, $\sigma_\phi$, is found following \citet{Naghizadeh1993}. We convert $\phi$ to the polarization angle with respect to the Galactic reference frame, $\theta$, following \cite{erratum}. Angles in the paper conform to the IAU convention.

\subsection{Ancillary data}

We use the \textit{Planck} maps of Stokes I, Q, U at 353 GHz that have been processed with the Generalized Needlet Internal Linear Combination (GNILC) algorithm to filter out Cosmic Infrared Background anisotropies \citep{Planck_IV_2018}. We follow the procedures outlined in \cite{Planck_2018_XII} to construct maps of the fractional linear polarization $p_{353} = P_{353}/I$, the polarized intensity $P_{353} = \sqrt{Q^2+U^2}$ and the polarization angle $\chi_{353} = 0.5\arctan(-U/Q)$. The maps have a resolution (FWHM) of 80$\arcmin$ and are sampled using the Hierarchical Equal Area iso-Latitude Pixelization scheme \citep[HEALPix][]{gorski2005} at a resolution \texttt{nside} 2048, which we downgrade to \texttt{nside} 128 to avoid oversampling.

To estimate the reddening, $E(B-V)$, and its uncertainty, we make use of four different reddening maps: 
\begin{itemize}
\item[1.] The GNILC map of dust optical depth, $\tau_{353}$, \citep{PlanckGNILC2016} (PG). We adopt the conversion $E(B-V) = (1.49\times 10^4$ mag)$\cdot \, \tau_{353}$ \citep{Planck_XI_2014}. The map beam size varies at high galactic latitude, but is generally $\sim$ 5$\arcmin$ at intermediate latitude.
\item[2.] The $A_V$ map from \citet{planckxxix} (PDL), which was created by fitting the dust model of \citet{DraineLi2007} to dust emission and renormalized to match quasar extinctions. The beam size is 5$\arcmin$. To convert to reddening we assume a ratio of total to selective extinction $R_V$ = 3.1.
\item[3.] The reddening map of \citet{SFD1998}(SFD), which is based primarily on 100 $\mu$m emission as measured by IRAS at 6$\arcmin$ resolution. 
\item[4.] The line-of-sight integrated map of reddening from \citet{green2018} (G18)\footnote{From \url{https://lambda.gsfc.nasa.gov/product/foreground}}, which is based on stellar photometry. In the selected region the resolution of this map is 7\arcmin.
\end{itemize}

\begin{figure*}[ht!]
\centering
\includegraphics[scale=1]{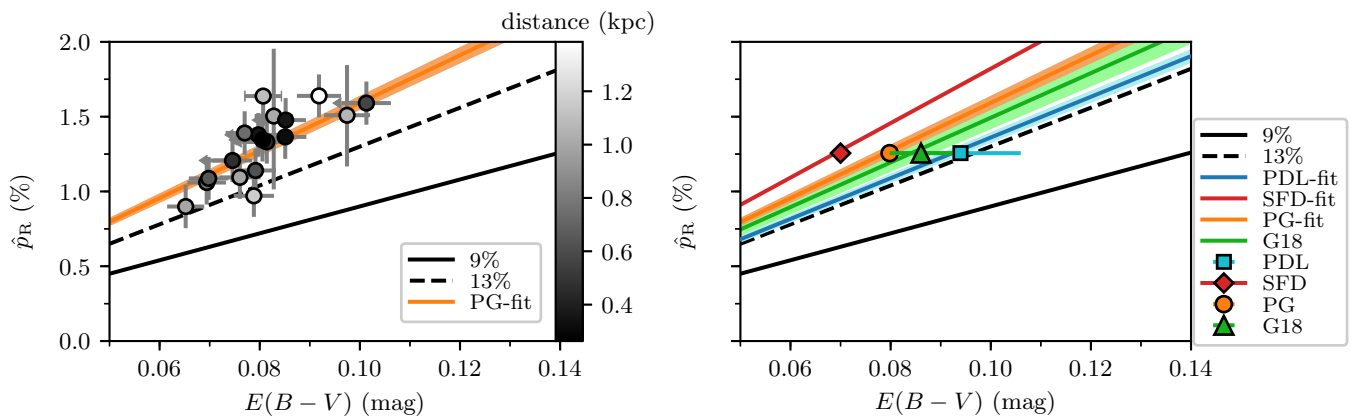}
\caption{Left: $\hat{p}_{\rm R}$ versus the PG estimate of $E(B-V)$ for stars farther than 250\,pc (circles). Values of $\rm [p_R/E(B-V)]_{max}$ equal to 9\% and 13\% are indicated with black the solid and dashed lines, respectively. The linear fit to the data, has a slope of 15.9$\pm$0.4 \%\,mag$^{-1}$ (orange line and 1$\sigma$ uncertainty as shaded region). Each star is colored according to its distance (colorbar). Right: Effect of uncertainty in $E(B-V)$. A linear fit to the data from each reddening map (PG, orange, SFD, red, PDL, cyan, G18, green) is shown, with shaded regions marking 1$\sigma$ uncertainty on the slope. The weighted mean $\hat{p}_{\rm R}$ and $E(B-V)$ is shown for the $E(B-V)$ from PG (orange circle), SFD (red diamond), PDL (cyan square), and G18 (green triangle).}
\label{fig:pebv}
\end{figure*}

Both the SFD and G18 maps are on the same reddening scale, which must be recalibrated to match $E(B-V)$ from the other estimates. We follow \citet{SchlaflyFinkbeiner} to find  $E(B-V)$ = $E(B-V)_{\rm SFD}\times0.884$. We note that reddening maps are constructed by making an assumption about the reddening law\footnote{\url{http://argonaut.skymaps.info/usage}}. The uncertainty in the reddening law alone translates into a systematic uncertainty of $\sim$13\% for all maps used here. Additional systematic uncertainties arising, for example, from the conversion from far-infrared opacity to extinction \citep[see e.g.][]{fanciullo2015} affect maps based on far-infrared emission (PG, PDL, SFD).

Finally, we use data from the Galactic L-band Feed Array HI survey (GALFA-HI), Data release 2 \citep{GALFADR2}, to obtain the HI column density and the orientation of linear structures seen in HI emission.

\section{Results} \label{sec:results}

We investigate the $p_{353}$ in the selected area in Fig. \ref{fig:planckp} (left). The value of $p_{353}$ in this region is 22\%, equivalent to the 99.5 percentile of the distribution of $p_{353}$ throughout the sky. The value of $p_{353}$ ranges from 21-23\% for the range of Galactic offset values used in \cite{Planck_2018_XII} ($23-103$\,$\mu {\rm K}_{\rm CMB}$). 

Our observations in the R band yielded 17 significant detections of the fractional linear polarization ($\hat{p}_{\rm R}/\sigma_p \geq 3$) and 5 upper limits\footnote{The data will become available at http://cdsarc.u-strasbg.fr.}. The mean SNR in $\hat{p}_{\rm R}$ is 7 with values ranging from 1.5 to 12.
Figure~\ref{fig:planckp} (right) shows the positions of the stars on the PG map of total reddening along the line-of-sight. The stars are located in areas with $E(B-V)$ ranging from 0.065$\pm0.003$ mag to 0.104$\pm0.005$ mag. A segment at the position of each star shows the orientation of the measured linear polarization. The distribution of $\theta$ (including only significant detections) has a mean of 109$^\circ$ and a standard deviation of 5$^\circ$.

In Appendix \ref{ssec:reddening}, we examine the distribution of dust along the line of sight and find that the dominant contribution to the observed starlight polarization comes from a cloud located at 250 pc. In the following, we do not consider stars nearer than 250 pc as they are foreground and have negligible polarization.

\subsection{$p/E(B-V)$}
\label{ssec:pebv}

The correlation of $\hat{p}_{\rm R}$ with $E(B-V)_{\rm PG}$ for stars farther than 250\,pc is shown in Figure \ref{fig:pebv} (left). All of the measurements fall above the $p_R$ = 9\% $E(B-V)$\,${\rm mag^{-1}}$ limit, consistent with the prediction that high levels of $p_{353}$ correspond to high levels of $p/E(B-V)$. The best-fit line with zero intercept has a slope of 15.9$\pm$0.4\%\,mag$^{-1}$. 
We note that $p_R$ and $p_V$ are expected to differ by only a few percent for typical Serkowski Law parameters (see Equation~\ref{eqn:serkowski}), and indeed $p_R$ and $p_V$ are consistent to within our measurement uncertainties for all stars with measurements of both (see Section~\ref{subsec:serkowski}). We therefore do not differentiate between $p_R/E(B-V)$ and $p_V/E(B-V)$.

The choice of reddening map has a significant impact on the best-fit slope (Fig. \ref{fig:pebv}, right). We find values ranging from 13.6$\pm$0.3\%\,mag$^{-1}$ (PDL) to 18.2$\pm$0.5\%\,mag$^{-1}$ (SFD). This variation reflects the unaccounted for systematic uncertainties in each of the reddening maps (see Section \ref{sec:obs}). Despite the uncertainties, all reddening estimates reject 9\%\,mag$^{-1}$ as an upper limit. 

A potential systematic effect arises from the coarse angular resolution of the each reddening map compared to the scales probed by stellar measurements. We could be underestimating the true reddening if clumpy regions of high reddening are smeared out by the beam \citep[e.g.,][]{Pereyra2007}. To account for such an effect, we calculate the average $E(B-V)$ in the observed 80\arcmin\ region and compare it to the mean $\hat{p}_{\rm R}$ (found through the weighted mean $q$ and $u$). The ratio of mean $\hat{p}_{\rm R}$ to mean $E(B-V)$ is 15.7$\pm$0.5\%\,${\rm mag^{-1}}$ (PG), 13.4$\pm$1.8\%\,${\rm mag^{-1}}$ (PDL), 17.9$\pm$0.5\%\,${\rm mag^{-1}}$ (SFD), 14.6$\pm$1.1\%\,${\rm mag^{-1}}$ (G18). 

Our estimates of $p_V/E(B-V)$ appear to be higher than 13\%\,mag$^{-1}$, inferred by \citet{Planck_2018_XII}. There are a few considerations that must be taken into account for a comparison to be made. First, the derivation of $\left[p_V/E(B-V)\right]_{\rm max}$ = 13\%\,mag$^{-1}$ was based on a slightly lower $p_{353}$ (20\%) than found in this region. This would allow for slightly higher values of $p_V/E(B-V)$. Second, since we do not subtract possible background reddening for stars that are nearer than 600 pc (where the dust column saturates, see Appendix \ref{ssec:reddening}), we are overestimating the reddening towards these stars and therefore our derivations of $p_V/E(B-V)$ are lower limits. By taking into account this fact, and the systematic uncertainty in the reddening law (Section \ref{sec:obs}), we can place a conservative lower limit on $\left[p_V/E(B-V)\right]_{\rm max}$ of 13\%\,${\rm mag^{-1}}$ (G18).

It is interesting to compare the average emission-to-extinction polarization ratios $R_{P/p} = P_{353}/\hat{p}_{\rm R}$ and $R_{S/V} = 3.1/1.086 \, p_{353}/[p_V/E(B-V)]$ in this region to those found over the larger sky area used in \citet{Planck_2018_XII}. As we only have one independent measurement of $P_{353}$ in the region, we calculate $R_{P/p}$ as the ratio of $P_{353}$ to the mean $\hat{p}_{\rm R}$ (and similarly for $R_{S/V}$). The value of $R_{S/V}$ inherits the large uncertainties from the determination of $E(B-V)$. We find $R_{S/V}$ of 3.9$\pm$0.1 (PG), 4.6$\pm$0.6 (PDL), 3.4$\pm$0.1 (SFD), 4.2$\pm$0.3 (G18). This is consistent with 4.2$\pm$0.5
found in \citet{Planck_2018_XII}.
In contrast, we can place much stronger constraints on our estimate of $R_{P/p}$. We find this ratio to be $4.1\pm0.1$\,MJy\,sr$^{-1}$, lower than that of $\sim$5.4 found in \citet{Planck_2018_XII} for the hydrogen column density in this region. 
Further investigation is required to understand the cause of this difference.

\subsection{Wavelength dependence of polarization}\label{subsec:serkowski}

The wavelength dependence of the optical polarized extinction $p_{\rm opt}$ is often parameterized as (Serkowski et al 1975):
\begin{equation}
p_{\rm opt}(\lambda) = p_{\rm max} \, e^{-K \ln^2(\lambda/\lambda_{\rm max})},
\label{eqn:serkowski}
\end{equation}
where $p_{\rm opt}$ has a maximum value of $p_{\rm max}$ at wavelength $\lambda_{\rm max}$ and $K$ is a parameter governing the width of the profile.

\begin{figure}[ht!]
\centering
\includegraphics[scale=1]{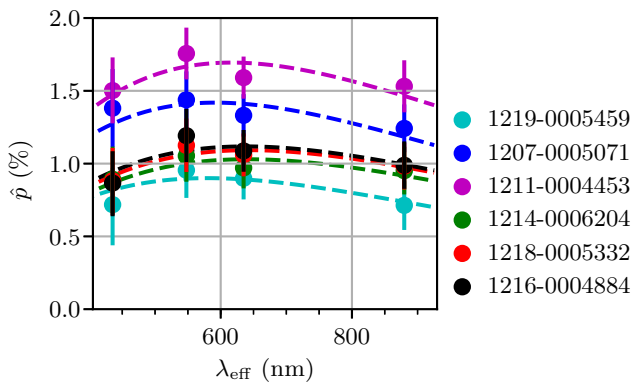}
\caption{Wavelength dependence of polarization for 6 of the stars in the R-band sample (USNO-B1 identifiers shown in the legend). Debiased fractional linear polarization ($\hat{p}$) measured in the Johnsons-Cousins B, V, R and I bands (horizontal axis shows effective wavelength). For each star, a line shows the best-fit Serkowski law (Equation~\ref{eqn:serkowski}, with $K$ set to 1.15).}
\label{fig:serkowski}
\end{figure}

Figure~\ref{fig:serkowski} shows the measured fractional linear polarization as a function of wavelength for the 6 stars measured in the B, V, R and I bands. Fixing $K = 1.15$ \citep{Serkowski1975} and leaving $\lambda_{\rm max}$ and $p_{\rm max}$ as free parameters, we fit the six polarization profiles to Equation~\ref{eqn:serkowski} using weighted least-squares.

\begin{figure*}[ht!]
\centering
\includegraphics[scale=1]{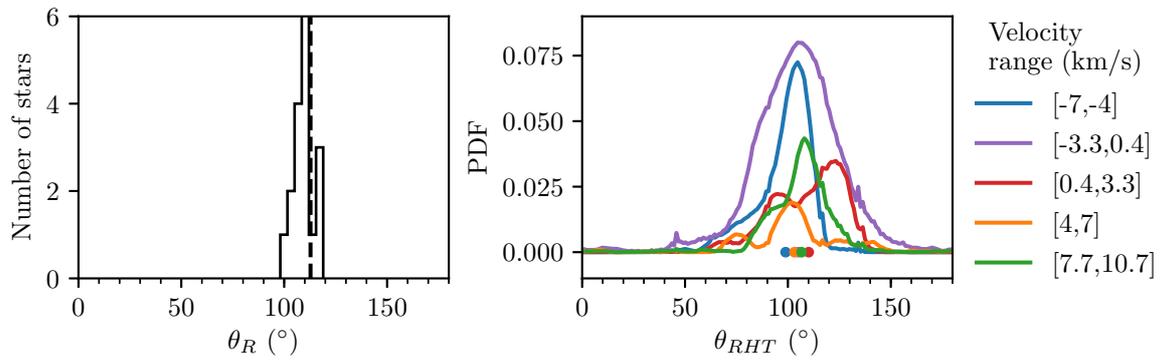}
\caption{Left: Distribution of stellar polarization angles, $\theta_R$ for stars farther than 250\,pc (solid line). The dashed vertical line shows the $\chi_{353}$+90$^\circ$ from \textit{Planck} at 80\arcmin resolution. Right: Probability density functions of the orientation of linear HI structures ($\theta_{RHT}$) for different velocity ranges (shown on the right). We only consider velocity ranges within the peak of the HI intensity spectrum. The filled circles in the bottom mark the mean value, $\left \langle \theta_{RHT}\right\rangle$ for each velocity range. All angles are measured in the Galactic reference frame, according to the IAU convention.}
\label{fig:anglesrht}
\end{figure*}

The fit $\lambda_{\rm max}$ are in the range $\rm 580 \pm 46\,nm$ to $\rm 643 \pm 34\,nm$, while the fit $p_{\rm max}$ are in the range $0.9\%-1.7\%$. Although limited by the uncertainties of the measurements, we find no evidence for substantial differences between the best-fit value of $\lambda_{\rm max}$ in this sample compared to canonical values \citep{Serkowski1975}. If grain alignment were unusually efficient in this region with smaller grains able to be aligned, we might have expected a shift of $\lambda_{\rm max}$ toward shorter wavelengths than the typical 550\,nm \citep{Martin1999}. This is not observed.

\subsection{Magnetic field morphology}
\label{ssec:angles}
If the magnetic field morphology is responsible for the high submillimeter polarization fractions, then any region with high $p_{353}$ should have a magnetic field that lies mostly in the plane of the sky with no substantial change in orientation either along the line of sight or within the beam.

The orientation of the magnetic field as projected on the plane of the sky is probed by the optical polarization angle $\theta_R$ as well as the submillimeter polarization angle rotated by 90$^\circ$. Within an $80\arcmin$ beam centered on this region we find $\chi_{353}+90^\circ=112.8\pm0.6^\circ$. The polarization angles inferred from stellar polarimery do not vary significantly across the region (Figure \ref{fig:anglesrht}, left). The standard deviation of the distribution of $\theta_R$ is consistent with arising from measurement uncertainties, suggesting little beam depolarization.

Linear structures seen in HI have also been shown to trace the interstellar magnetic field \citep{Clark2014}. Recently, \citet{Clark2018} proposed that changes in the orientation of these structures in velocity space can be used to probe variations of the magnetic field orientation along the line of sight. 
By applying the Rolling Hough Transform (RHT) to maps of the HI emission as in \citet{Clark2014}, \citet{GALFADR2} quantify the orientation of HI structures for different velocity channels for the entire GALFA-HI footprint. Figure~\ref{fig:anglesrht} illustrates the probability density function of the orientation angle $\theta_{\rm RHT}$ of linear HI structures in our selected region for different velocity channels. 
The mean $\theta_{\rm RHT}$ cover a small range (99$^\circ - 107^\circ$), as expected if the magnetic field exhibits minimal variation along the sightline.

\section{Concluding remarks}
\label{sec:summary}

We have investigated the properties of starlight polarization in a region where $p_{353}\simeq 20\%$, with the goal of understanding the origin of these high levels of polarization. We have found that stars tracing the entire dust column show high $p_V/E(B-V)$, significantly exceeding the classical upper limit of 9\%\,mag$^{-1}$. The stellar polarimetry and the HI data suggest a well-ordered magnetic field both across the region and along the line of sight. The wavelength dependence of the optical polarization is consistent with a standard Serkowski Law with no shift in $\lambda_{\rm max}$ to shorter wavelengths as might be expected if efficient grain alignment extended to unusually small grain sizes.


Although the studied region has unusually high $p_{353}$, it appears typical in both the wavelength dependence of the optical polarization and in the ratio of polarized emission to polarized extinction. Thus, there is no clear indication that the high polarization fractions are attributable to dust properties such as enhanced alignment or unusual composition. Rather, our analysis favors a scenario in which the observed high $p_{353}$ and $p_V/E(B-V)$ arise simply from a favorable magnetic field geometry. This extends the conclusion of \citet{Planck_Int_XX} that the the magnetic field morphology can account for the observed polarized emission properties, even in the highest $p_{353}$ regions of the sky. 

Our findings show that the 13\%\,mag$^{-1}$ inferred by \citet{Planck_2018_XII} is likely a lower limit on the true $\left[p_V/E(B-V)\right]_{\rm max}$, confirming that the classical upper limit of $p_V/E(B-V) \leq 9\%\,{\rm mag}^{-1}$ should be revised. Further progress calls for more accurate determination of stellar $E(B-V)$.

Finally, we note that the observed region is not unique in its unusually high $p_V/E(B-V)$, with other highly polarized stars reported elsewhere in the literature \citep[e.g.,][]{Whittet1994,Pereyra2007,Andersson2007,Frisch2015,Panopoulou2015,skalidis,Planck_2018_XII,Panopoulou2019}. However, as discussed in Section~\ref{ssec:pebv}, the determination of $p_V/E(B-V)$ is limited in large part by accurate determination of interstellar reddening. For instance, \citet{Gontcharov2019} find that the same sample of nearby stars may reject the classical $[p_V/E(B-V)]_{\rm max}$ limit or not, depending on the choice of reddening map.
Through polarimetry of many stars in a single region of high submillimeter polarization, we have shown that the mean $p_V/E(B-V)$ in the region is inconsistent with an upper limit of 9\%\,mag$^{-1}$ regardless of which reddening map is used.

\begin{acknowledgements}
      
We thank V. Guillet for his insightful review, V. Pelgrims, for helpful comments, and P. Martin, S. Clark and the ESA/Cosmos helpdesk for advice on using \textit{Planck} data. G.\,V.\,P. acknowledges support from the National Science Foundation, under grant number AST-1611547. R.\,S., D.\,B. and K.\,T. acknowledge support from the European Research Council under the European Union's Horizon 2020 research and innovation program, under grant agreement No 771282.

Based on observations obtained with \textit{Planck} (\url{http://www.esa.int/Planck}), a European Space Agency (ESA) science mission. This work has made use of data from the ESA mission
{\it Gaia} (\url{https://www.cosmos.esa.int/gaia}), processed by the {\it Gaia}
Data Processing and Analysis Consortium (DPAC,
\url{https://www.cosmos.esa.int/web/gaia/dpac/consortium}). Funding for the DPAC
has been provided by national institutions, in particular the institutions
participating in the {\it Gaia} Multilateral Agreement. 

Based on Galactic ALFA HI (GALFA HI) survey data obtained with the Arecibo L-band Feed Array (ALFA) on the Arecibo 305m telescope. The Arecibo Observatory is a facility of the National Science Foundation (NSF) operated by SRI International in alliance with the Universities Space Research Association (USRA) and UMET under a cooperative agreement. The GALFA HI surveys are funded by the NSF through grants to Columbia University, the University of Wisconsin, and the University of California.
\end{acknowledgements}

\appendix
\section{The distribution of dust along the line of sight}
\label{ssec:reddening}

\begin{figure}[ht!]
\centering
\includegraphics[scale=1]{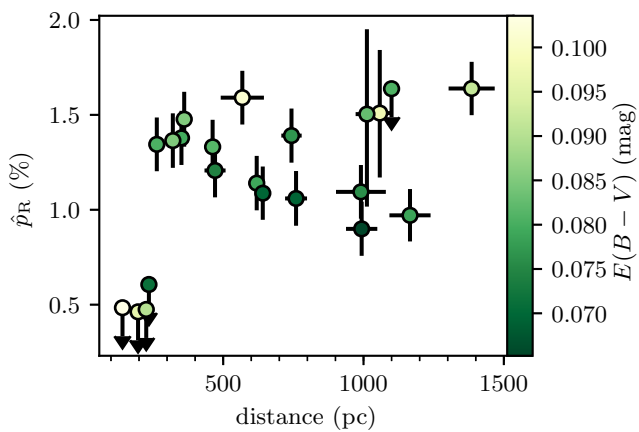}
\caption{Debiased fractional linear polarization of stars measured in the R band ($\hat{p}_{\rm R}$) versus stellar distance. Points are colored according to the $E(B-V)$ from PG at the position of the star. We only show 2$\sigma$ upper limits on $\hat{p}_{\rm R}$ for stars with $\hat{p}_{\rm R}/\sigma_p < 3$.}
\label{fig:pd}
\end{figure}

Figure~\ref{fig:pd} shows the dependence of $\hat{p}_{\rm R}$ with distance. Polarization is very low for the nearest 4 stars in our sample. For all stars within 250\,pc we have obtained only (2$\sigma$) upper limits on $\hat{p}_{\rm R}$ of 0.6\% or lower. At distances farther than that of the fourth star (235$\pm3$ pc) we find highly significant detections of $\hat{p}_{\rm R}$ with a mean of 1.3\%. This abrupt change strongly suggests that there is a dust component (cloud) located at a distance between that of the fourth and fifth nearest star. If we take into account the distance of the fifth nearest star ($263\pm5$\,pc) we can place conservative limits on the cloud distance of $232-268$\,pc. We adopt 250\,pc as our estimate of the distance to this cloud. 

We note that in Fig. \ref{fig:pd} some stars at large distances are less polarized than nearby stars. This can appear as a reduction of $\hat{p}_{\rm R}$ with distance as might be caused by additional dust components with different polarization orientations. However, the stars at greater distances lie along sightlines with lower $E(B-V)$ (colors in Fig.~\ref{fig:pd}), thus explaining the apparent trend.

We obtain a rough estimate of the distance at which the dust column ends from the 3D map of reddening of \cite{green2018}. For all sightlines, reddening saturates at a distance of $\sim 600$\,pc. This suggests that stars located farther than 600\,pc are tracing the entire dust column. For these stars the $E(B-V)$ from the maps presented in Section \ref{sec:obs} will likely reflect their true reddening.
For the 6 stars at distances between 250 pc and 600 pc, we do not correct for background reddening, which is within the G18 systematic uncertainty of 0.02 mag. Instead we assign the full line-of-sight reddening. This overestimates their reddening so that our estimates of $p/E(B-V)$ (Section \ref{ssec:pebv}) are conservative lower limits.

\bibliography{aanda}

\begin{thebibliography}{38}
\expandafter\ifx\csname natexlab\endcsname\relax\def\natexlab#1{#1}\fi

\bibitem[{{Andersson} \& {Potter}(2007)}]{Andersson2007}
{Andersson}, B.-G. \& {Potter}, S.~B. 2007, \apj, 665, 369

\bibitem[{{Bailer-Jones} {et~al.}(2018){Bailer-Jones}, {Rybizki}, {Fouesneau},
  {Mantelet}, \& {Andrae}}]{bailer-jones}
{Bailer-Jones}, C.~A.~L., {Rybizki}, J., {Fouesneau}, M., {Mantelet}, G., \&
  {Andrae}, R. 2018, \aj, 156, 58

\bibitem[{{Clark}(2018)}]{Clark2018}
{Clark}, S.~E. 2018, \apjl, 857, L10

\bibitem[{{Clark} {et~al.}(2014){Clark}, {Peek}, \& {Putman}}]{Clark2014}
{Clark}, S.~E., {Peek}, J.~E.~G., \& {Putman}, M.~E. 2014, \apj, 789, 82

\bibitem[{{Draine} \& {Fraisse}(2009)}]{Draine2009}
{Draine}, B.~T. \& {Fraisse}, A.~A. 2009, \apj, 696, 1

\bibitem[{{Draine} \& {Li}(2007)}]{DraineLi2007}
{Draine}, B.~T. \& {Li}, A. 2007, \apj, 657, 810

\bibitem[{{Fanciullo} {et~al.}(2015){Fanciullo}, {Guillet}, {Aniano}, {Jones},
  {Ysard}, {Miville-Desch{\^e}nes}, {Boulanger}, \&
  {K{\"o}hler}}]{fanciullo2015}
{Fanciullo}, L., {Guillet}, V., {Aniano}, G., {et~al.} 2015, \aap, 580, A136

\bibitem[{{Frisch} {et~al.}(2015){Frisch}, {Berdyugin}, {Piirola}, {Magalhaes},
  {Seriacopi}, {Wiktorowicz}, {Andersson}, {Funsten}, {McComas}, {Schwadron},
  {Slavin}, {Hanson}, \& {Fu}}]{Frisch2015}
{Frisch}, P.~C., {Berdyugin}, A., {Piirola}, V., {et~al.} 2015, \apj, 814, 112

\bibitem[{{Gontcharov} \& {Mosenkov}(2019)}]{Gontcharov2019}
{Gontcharov}, G.~A. \& {Mosenkov}, A.~V. 2019, \mnras, 483, 299

\bibitem[{{G{\'o}rski} {et~al.}(2005){G{\'o}rski}, {Hivon}, {Banday},
  {Wandelt}, {Hansen}, {Reinecke}, \& {Bartelmann}}]{gorski2005}
{G{\'o}rski}, K.~M., {Hivon}, E., {Banday}, A.~J., {et~al.} 2005, \apj, 622,
  759

\bibitem[{{Green} {et~al.}(2018){Green}, {Schlafly}, {Finkbeiner}, {Rix},
  {Martin}, {Burgett}, {Draper}, {Flewelling}, {Hodapp}, {Kaiser}, {Kudritzki},
  {Magnier}, {Metcalfe}, {Tonry}, {Wainscoat}, \& {Waters}}]{green2018}
{Green}, G.~M., {Schlafly}, E.~F., {Finkbeiner}, D., {et~al.} 2018, \mnras,
  478, 651

\bibitem[{{Guillet} {et~al.}(2018){Guillet}, {Fanciullo}, {Verstraete},
  {Boulanger}, {Jones}, {Miville-Desch{\^e}nes}, {Ysard}, {Levrier}, \&
  {Alves}}]{Guillet2018}
{Guillet}, V., {Fanciullo}, L., {Verstraete}, L., {et~al.} 2018, \aap, 610, A16

\bibitem[{{Hiltner}(1956)}]{Hiltner1956}
{Hiltner}, W.~A. 1956, \apjs, 2, 389

\bibitem[{{Martin} {et~al.}(1999){Martin}, {Clayton}, \& {Wolff}}]{Martin1999}
{Martin}, P.~G., {Clayton}, G.~C., \& {Wolff}, M.~J. 1999, \apj, 510, 905

\bibitem[{{Monet} {et~al.}(2003){Monet}, {Levine}, {Canzian}, {Ables}, {Bird},
  {Dahn}, {Guetter}, {Harris}, {Henden}, {Leggett}, {Levison}, {Luginbuhl},
  {Martini}, {Monet}, {Munn}, {Pier}, {Rhodes}, {Riepe}, {Sell}, {Stone},
  {Vrba}, {Walker}, {Westerhout}, {Brucato}, {Reid}, {Schoening}, {Hartley},
  {Read}, \& {Tritton}}]{usnob}
{Monet}, D.~G., {Levine}, S.~E., {Canzian}, B., {et~al.} 2003, \aj, 125, 984

\bibitem[{{Naghizadeh-Khouei} \& {Clarke}(1993)}]{Naghizadeh1993}
{Naghizadeh-Khouei}, J. \& {Clarke}, D. 1993, \aap, 274, 968

\bibitem[{{Panopoulou} {et~al.}(2016){Panopoulou}, {Tassis}, {Blinov},
  {Pavlidou}, {King}, {Paleologou}, {Ramaprakash}, {Angelakis},
  {Balokovi{\'c}}, {Das}, {Feiler}, {Hovatta}, {Khodade}, {Kiehlmann}, {Kus},
  {Kylafis}, {Liodakis}, {Mahabal}, {Modi}, {Myserlis}, {Papadakis},
  {Papamastorakis}, {Pazderska}, {Pazderski}, {Pearson}, {Rajarshi},
  {Readhead}, {Reig}, \& {Zensus}}]{erratum}
{Panopoulou}, G., {Tassis}, K., {Blinov}, D., {et~al.} 2016, \mnras, 462, 2011

\bibitem[{{Panopoulou} {et~al.}(2015){Panopoulou}, {Tassis}, {Blinov},
  {Pavlidou}, {King}, {Paleologou}, {Ramaprakash}, {Angelakis},
  {Balokovi{\'c}}, {Das}, {Feiler}, {Hovatta}, {Khodade}, {Kiehlmann}, {Kus},
  {Kylafis}, {Liodakis}, {Mahabal}, {Modi}, {Myserlis}, {Papadakis},
  {Papamastorakis}, {Pazderska}, {Pazderski}, {Pearson}, {Rajarshi},
  {Readhead}, {Reig}, \& {Zensus}}]{Panopoulou2015}
{Panopoulou}, G., {Tassis}, K., {Blinov}, D., {et~al.} 2015, \mnras, 452, 715

\bibitem[{{Panopoulou} {et~al.}(2019){Panopoulou}, {Tassis}, {Skalidis},
  {Blinov}, {Liodakis}, {Pavlidou}, {Potter}, {Ramaprakash}, {Readhead}, \&
  {Wehus}}]{Panopoulou2019}
{Panopoulou}, G.~V., {Tassis}, K., {Skalidis}, R., {et~al.} 2019, \apj, 872, 56

\bibitem[{{Peek} {et~al.}(2018){Peek}, {Babler}, {Zheng}, {Clark}, {Douglas},
  {Korpela}, {Putman}, {Stanimirovi{\'c}}, {Gibson}, \& {Heiles}}]{GALFADR2}
{Peek}, J.~E.~G., {Babler}, B.~L., {Zheng}, Y., {et~al.} 2018, \apjs, 234, 2

\bibitem[{{Pereyra} \& {Magalh{\~a}es}(2007)}]{Pereyra2007}
{Pereyra}, A. \& {Magalh{\~a}es}, A.~M. 2007, \apj, 662, 1014

\bibitem[{{Planck Collaboration Int.~XIX}(2015)}]{Planck_Int_XIX}
{Planck Collaboration Int.~XIX}. 2015, \aap, 576, A104

\bibitem[{{Planck Collaboration Int. XLVIII}(2016)}]{PlanckGNILC2016}
{Planck Collaboration Int. XLVIII}. 2016, \aap, 596, A109

\bibitem[{{Planck Collaboration Int.~XX}(2015)}]{Planck_Int_XX}
{Planck Collaboration Int.~XX}. 2015, \aap, 576, A105

\bibitem[{{Planck Collaboration Int.~XXI}(2015)}]{Planck_Int_XXI}
{Planck Collaboration Int.~XXI}. 2015, \aap, 576, A106

\bibitem[{{Planck Collaboration Int.~XXIX}(2016)}]{planckxxix}
{Planck Collaboration Int.~XXIX}. 2016, \aap, 586, A132

\bibitem[{{Planck Collaboration IV}(2018)}]{Planck_IV_2018}
{Planck Collaboration IV}. 2018, arXiv e-prints, arXiv:1807.06208

\bibitem[{{Planck Collaboration X}(2016)}]{Planck_2015_X}
{Planck Collaboration X}. 2016, \aap, 594, A10

\bibitem[{{Planck Collaboration XI}(2014)}]{Planck_XI_2014}
{Planck Collaboration XI}. 2014, \aap, 571, A11

\bibitem[{{Planck Collaboration XII}(2018)}]{Planck_2018_XII}
{Planck Collaboration XII}. 2018, ArXiv e-prints [\eprint[arXiv]{1807.06212}]

\bibitem[{{Plaszczynski} {et~al.}(2014){Plaszczynski}, {Montier}, {Levrier}, \&
  {Tristram}}]{plaszczynski}
{Plaszczynski}, S., {Montier}, L., {Levrier}, F., \& {Tristram}, M. 2014,
  \mnras, 439, 4048

\bibitem[{{Ramaprakash} {et~al.}(2019){Ramaprakash}, {Rajarshi}, {Das},
  {Khodade}, {Modi}, {Panopoulou}, {Maharana}, {Blinov}, {Angelakis},
  {Casadio}, {Fuhrmann}, {Hovatta}, {Kiehlmann}, {King}, {Kylafis},
  {Kougentakis}, {Kus}, {Mahabal}, {Marecki}, {Myserlis}, {Paterakis},
  {Paleologou}, {Liodakis}, {Papadakis}, {Papamastorakis}, {Pavlidou},
  {Pazderski}, {Pearson}, {Readhead}, {Reig}, {S{\l}owikowska}, {Tassis}, \&
  {Zensus}}]{ramaprakash}
{Ramaprakash}, A.~N., {Rajarshi}, C.~V., {Das}, H.~K., {et~al.} 2019, \mnras
  [\eprint[arXiv]{1902.08367}]

\bibitem[{{Schlafly} \& {Finkbeiner}(2011)}]{SchlaflyFinkbeiner}
{Schlafly}, E.~F. \& {Finkbeiner}, D.~P. 2011, \apj, 737, 103

\bibitem[{{Schlegel} {et~al.}(1998){Schlegel}, {Finkbeiner}, \&
  {Davis}}]{SFD1998}
{Schlegel}, D.~J., {Finkbeiner}, D.~P., \& {Davis}, M. 1998, \apj, 500, 525

\bibitem[{{Schmidt} {et~al.}(1992){Schmidt}, {Elston}, \& {Lupie}}]{schmidt}
{Schmidt}, G.~D., {Elston}, R., \& {Lupie}, O.~L. 1992, \aj, 104, 1563

\bibitem[{{Serkowski} {et~al.}(1975){Serkowski}, {Mathewson}, \&
  {Ford}}]{Serkowski1975}
{Serkowski}, K., {Mathewson}, D.~S., \& {Ford}, V.~L. 1975, \apj, 196, 261

\bibitem[{{Skalidis} {et~al.}(2018){Skalidis}, {Panopoulou}, {Tassis},
  {Pavlidou}, {Blinov}, {Komis}, \& {Liodakis}}]{skalidis}
{Skalidis}, R., {Panopoulou}, G.~V., {Tassis}, K., {et~al.} 2018, \aap, 616,
  A52

\bibitem[{{Whittet} {et~al.}(1994){Whittet}, {Gerakines}, {Carkner}, {Hough},
  {Martin}, {Prusti}, \& {Kilkenny}}]{Whittet1994}
{Whittet}, D.~C.~B., {Gerakines}, P.~A., {Carkner}, A.~L., {et~al.} 1994,
  \mnras, 268, 1

\end{thebibliography}
\bibliographystyle{aa}

\end{document}